 \documentclass[final,5p,times,twocolumn]{elsarticle}

\usepackage{amssymb}
 \usepackage{lineno}

\usepackage{graphicx}

\journal{Nuclear Instruments and Methods in Physics Research Section A}

\begin{document}

\begin{frontmatter}



\title{Scintillation and Ionization Ratio of Liquid Argon for Electronic and Nuclear Recoils at Drift-Fields up to 3~kV/cm}


%\author{}
\author{T.Washimi\corref{cor1}}
\ead{washimi@kylab.sci.waseda.ac.jp}
\cortext[cor1]{Corresponding author.}

\author{M.Kimura}

\author{M.Tanaka}

\author{K.Yorita}
\ead{kohei.yorita@waseda.jp}

%\address{}
\address{Waseda University, Tokyo, Japan}

%\linenumbers

\begin{abstract}
A two-phase argon detector has high discrimination power between electron recoil and nuclear recoil events 
based on the pulse shape discrimination and the ionization/scintillation ratio (S2/S1).
This character is very suitable for the dark matter search to establish the low background experiment.
However, the basic properties of S2/S1 of argon are not well known, as compared with xenon.
We report the evaluation of S2/S1 properties with a two-phase detector at drift-fields of 0.2--3.0~kV/cm.  
Finally, the discrimination power against electron recoil background of S2/S1 is discussed.
\end{abstract}


\end{frontmatter}

% \linenumbers


%\newpage
\section{Introduction} \label{sec:Intro}
Two-phase noble gas detector technology has been used widely for weakly interacting massive particle (WIMP) dark matter detection experiments 
(e.g. DarkSide-50~\cite{agnes2018darkside, agnes2018low}, LUX~\cite{akerib2017results}, PandaX-II~\cite{cui2017dark}, and XENON-1T~\cite{aprile2017first}). 
Its technology aims for electron recoil (ER) background rejection from nuclear recoil (NR) signal using ionization(S2)/scintillation(S1) ratio. 
However, DarkSide-50 does not make use of the S2/S1 ratio for background rejection.
It is well known that the S1 and S2 light yields depend on the strength of electric field, 
imposed in drift interaction region, mainly due to recombination effect of ionizing electrons.  
Such properties are well measured by previous experiments, such as  
SCENE~\cite{cao2015measurement} (0--0.97~kV/cm, 10.3--57.3~$\rm keV_{nr}$, nr : nuclear recoil) and
ARIS~\cite{agnes2018measurement} (0--0.5~kV/cm, 7.1--117.8~$\rm keV_{nr}$) 
where drift-fields are lower than 1~kV/cm and the ER/NR discrimination power of S2/S1 is not explicitly described. 
In this paper, we focus on the drift-field dependence of S2/S1 properties up to 3.0~kV/cm.
Although liquid argon (LAr) scintillation has strong pulse shape discrimination (PSD) power~\cite{amaudruz2016measurement}, 
to simplify, S2/S1 discrimination power is separately discussed from PSD property in this paper.

%%%%%%%%%%%%%%%%%%%%%%%%%%%%%%%%%%%%%%%%%%%%%%%%%%%%%%%%%%%%%%%%%%%
%\newpage
\section{Experimental setup and basic performance} \label{sec:Setup}
\begin{figure}[ht]
\centering
\includegraphics[width=6.5cm,clip]{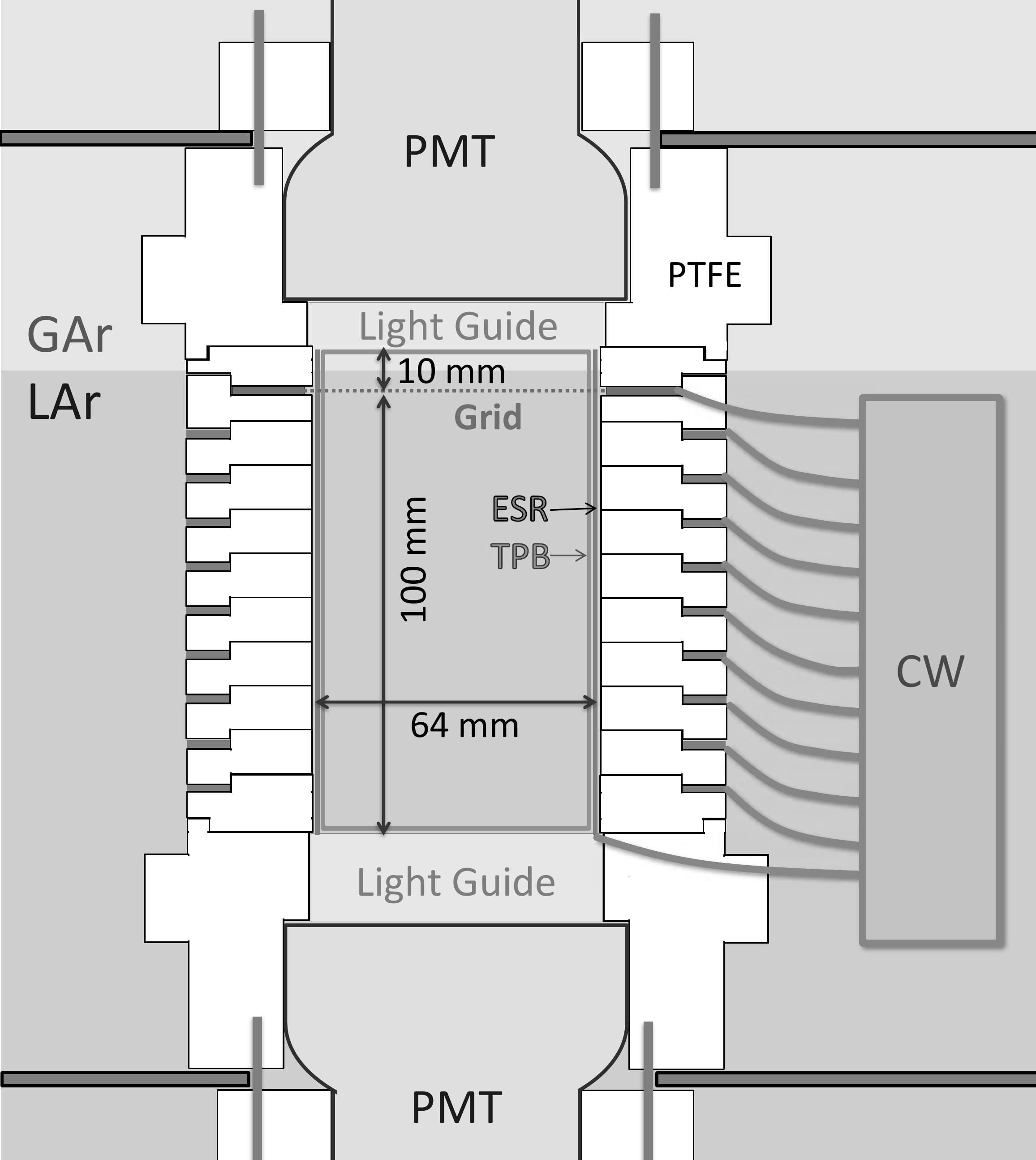}
\caption{Cross section of the detector}
\label{img:TPC}
\end{figure}

% teststand
This experiment was conducted in the Waseda liquid argon test stand~\cite{tanaka2013status, washimi2018study}. 
% TPC
Fig.~\ref{img:TPC} shows the schematic view of a two-phase detector we developed for this study. 
It mainly consists of a polytetrafluoroethylene (PTFE) cylinder with an active LAr volume of $\rm \phi 6.4~cm \times H 10~cm\ (\approx 0.5~kg)$. 
Two photomultiplier tubes (PMTs, HAMAMATSU R11065) are located on the top and bottom sides of the fiducial volume, 
where they are placed in contact with the transparent indium-tin-oxide (ITO) coated quartz light guides.
A stainless steel wire grid plane is inserted 1 cm below the top light guide. 
Tetraphenyl-butadiene (TPB) wavelength shifter (from ultra vacuum violet scintillation light to visible light) 
is deposited on the inner surfaces of the detector by vacuum evaporation method. 
The liquid argon surface is kept centered in height between the top light guide and the wire grid, 
and the operation inner gas pressure is kept at 1.5~atm stably.
% CW and E-Field
To form a high electric field time projection chamber (TPC),  a Cockcroft--Walton circuit (CW) generates high voltage (max: 30~kV) 
in the liquid argon and makes the drift-field (max:  3.0~kV/cm) in the detector. 
The potential difference of 4.5~kV is applied between the anode and the wire grid plane. 
By using the relative dielectric constant $\varepsilon$ and the position of liquid surface, the fields for electron extraction (in liquid, $\varepsilon=1.53$) 
and S2 emission (in gas, $\varepsilon=1.00$) are calculated to be 3.6~kV/cm and 5.4~kV/cm, respectively.

\begin{figure}[ht]
\centering
\includegraphics[width=8cm,clip]{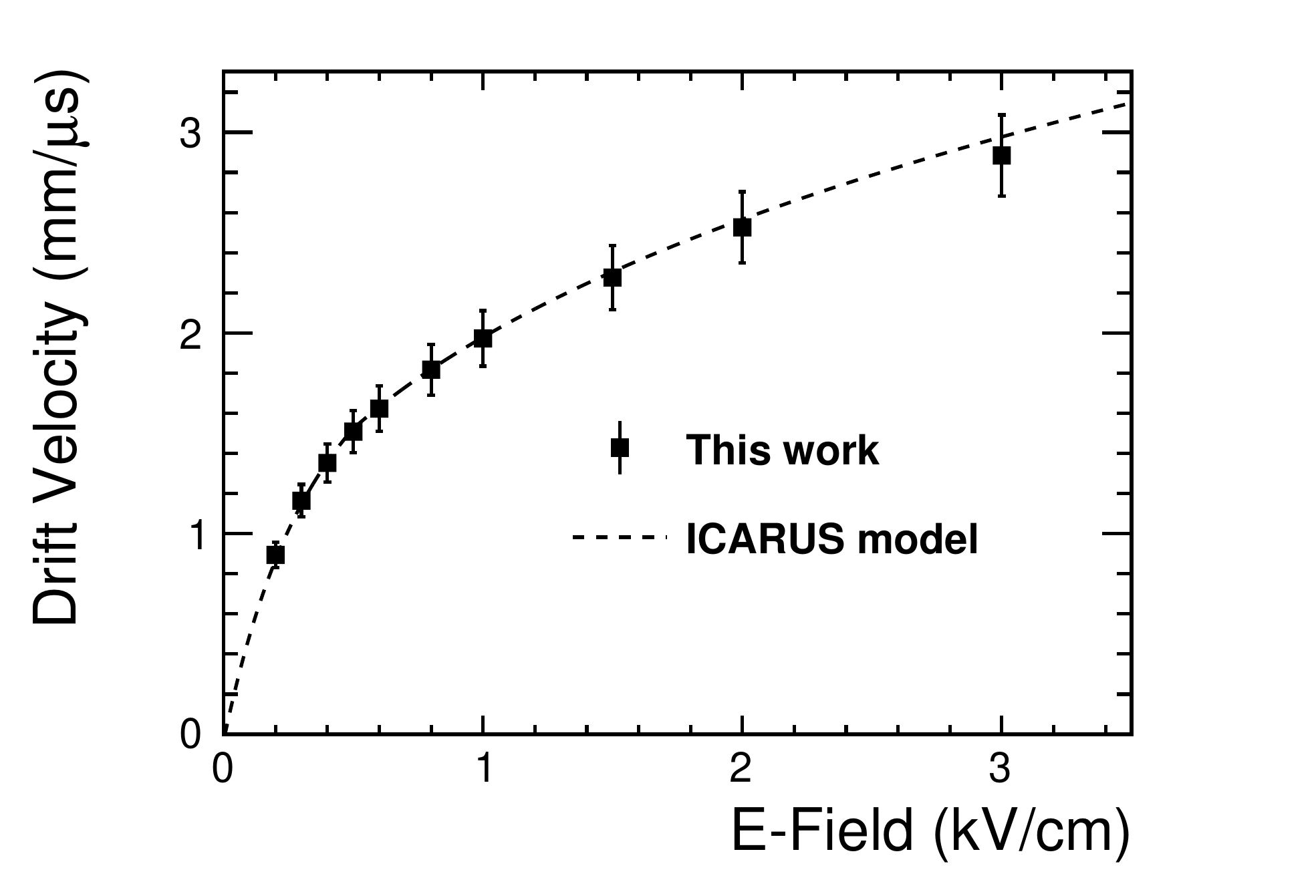}
\caption{Field dependence of drift velocity. 
The data points are our results, and the dashed line is calculated using model in the reference 
(ICARUS~\cite{amoruso2004analysis} and Walkowiak~\cite{walkowiak2000drift}).}
\label{img:DriftVelocity}
\end{figure}

%  RI source
For testing the system, $^{22}$Na and $^{252}$Cf radioactive sources are used for pure $\gamma$-ray (ER) events and neutron (NR) events, respectively.
These sources are located 1~m apart from the center of the TPC, outside of the chamber. 
To detect the associated $\gamma$-ray and determine the start time of flight (TOF), an NaI(Tl) scintillation counter is placed behind the source . 
In this setup, $\rm TOF = 3~ns$ for $\gamma$-ray and $\rm TOF = 50~ns$ for 2~MeV neutron.
% DAQ, trigger
The data acquisition system utilizes a 250~mega-samples per second flash ADC (SIS3316) with a three-channel coincidence 
trigger with the top PMT, the bottom PMT and the NaI(Tl) scintillator (coincidence width: 1~$\mu$s). 
% LY, purity
With this TPC configuration, the detection efficiency of S1 light is measured to be $\rm 5.7\pm 0.3~p.e./keV_{\rm ee}$ 
(ee : electron equivalent) for 511~keV $\gamma$-ray at null field, and the lifetime of the drift electron is measured to be  $1.9\pm 0.1$~ms.
Fig.~\ref{img:DriftVelocity} shows the drift velocity determined by using the collimated $^{22}$Na and $^{60}$Co $\gamma$-ray data, 
compared with a model from ICARUS~\cite{amoruso2004analysis} and Walkowiak~\cite{walkowiak2000drift}.

%%%%%%%%%%%%%%%%%%%%%%%%%%%%%%%%%%%%%%%%%%%%%%%%%%%%%%%%%%%%%%%%%%%%%%%%%%%%
%\newpage
\section{Measurements of ionization/scintillation ratio}
\label{sec:S2/S1_distribution}
\begin{figure}[htbp]
\centering
\includegraphics[width=9cm,clip]{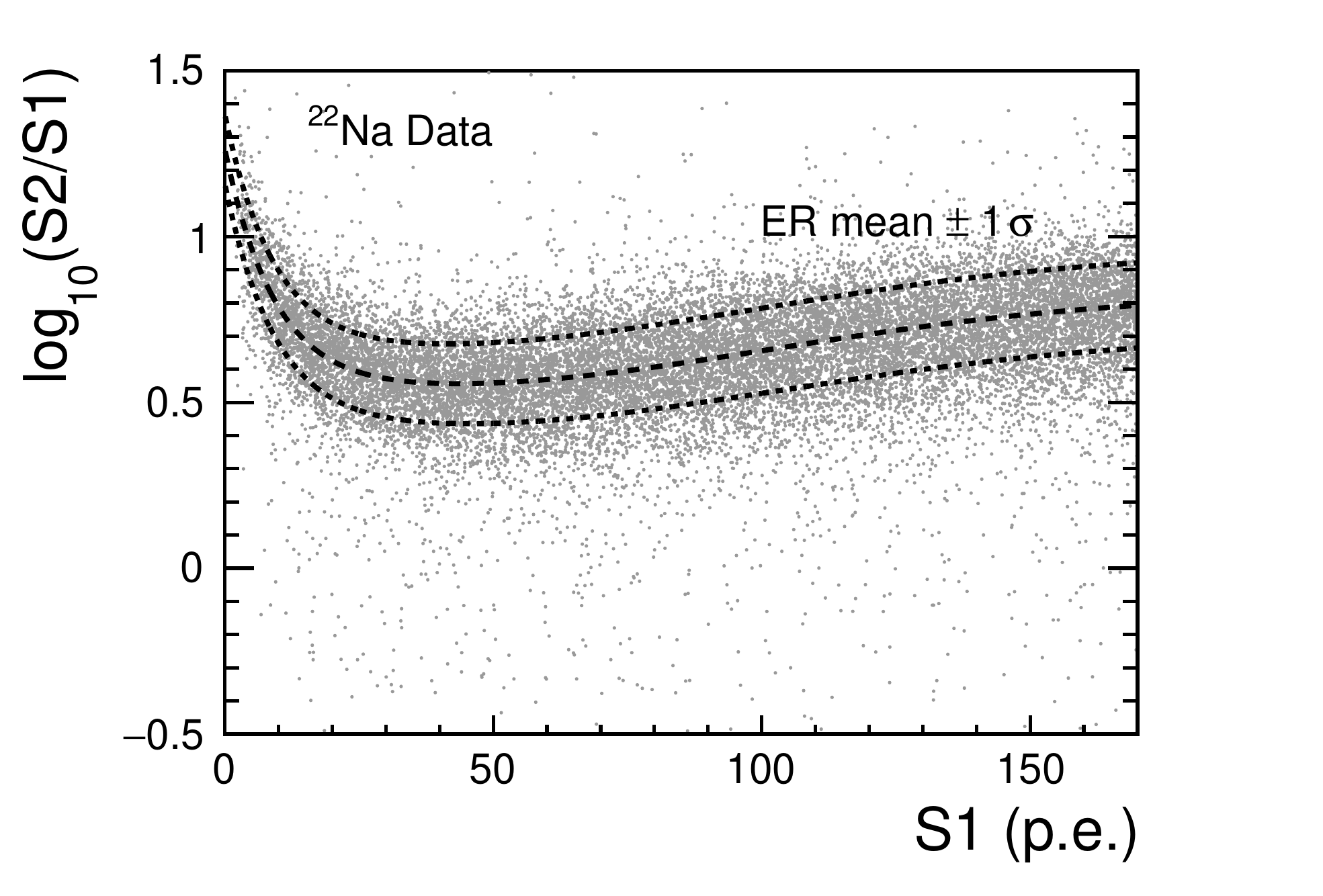}
\includegraphics[width=9cm,clip]{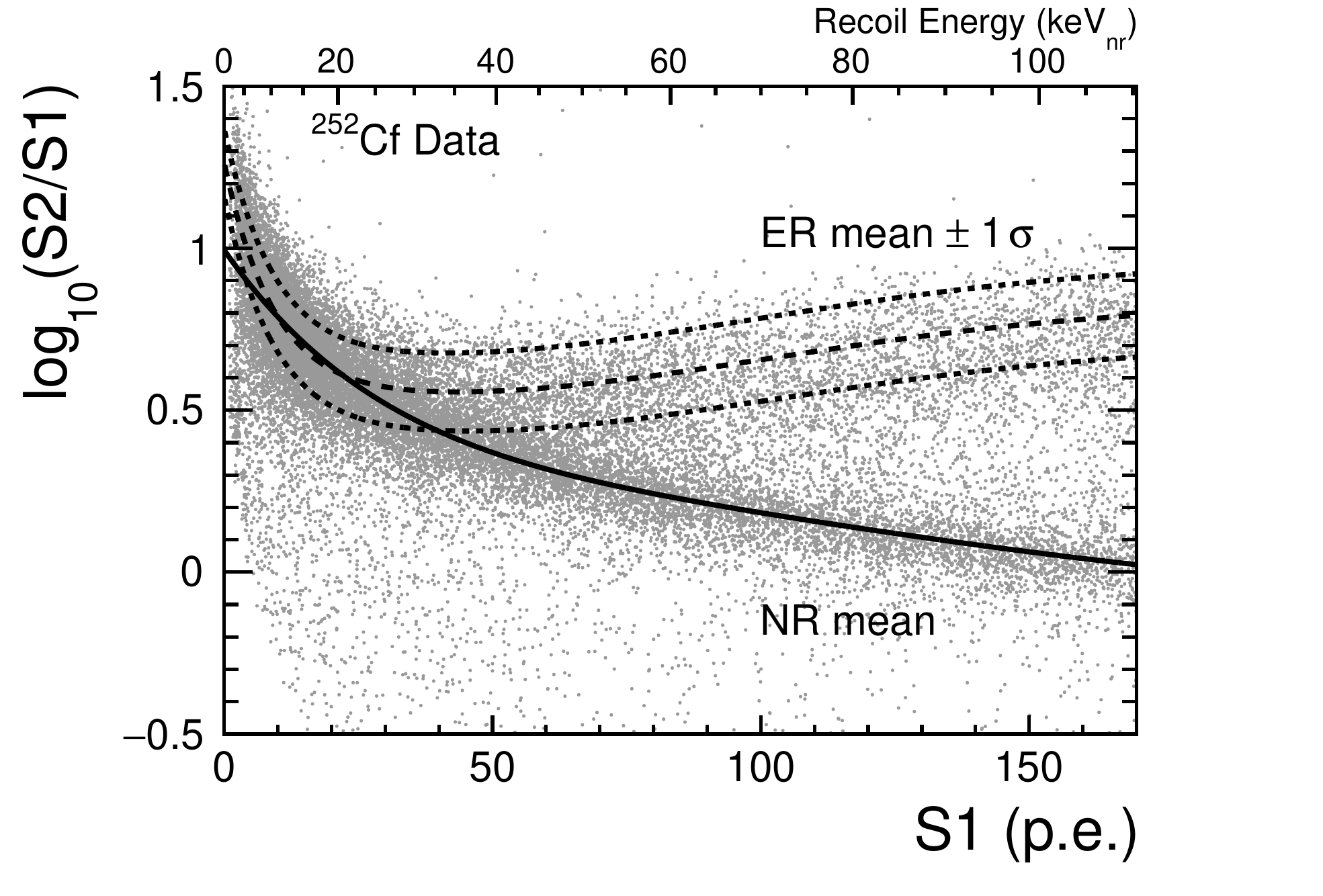}
\caption{$\log_{10}({\rm S2/S1})$ as a function of S1 light yield at the drift-field of 1~kV/cm. 
Top : $\rm ^{22}Na$ data,  Bottom : $\rm ^{252}Cf$ data.}
\label{img:S1-LogS2S1_1kV/cm}
\end{figure}

The upper plot in Fig.~\ref{img:S1-LogS2S1_1kV/cm} shows S2/S1 ratio ($\log_{10}({\rm S2/S1})$)
for pure ER events from $^{22}$Na source, as a function of S1 light yield at the drift-field of 1.0~kV/cm. 
The mean value ($\mu$) and 1$\sigma$ band are obtained by the Gaussian fit at each slice of S1 light yield.

The $^{252}$Cf data at 1.0~kV/cm, where neutron events are selected by using TOF information ($\rm TOF > 20~ns$),
is shown in the bottom plot of Fig.~\ref{img:S1-LogS2S1_1kV/cm}. 
The solid line is the mean($\mu$) of NR events, overlaid with a band of ER events from $^{22}$Na at the drift-field of 1.0~kV/cm.
Conversion calculation from S1 to recoil energy $E_{\rm nr}$ in the unit of $\rm keV_{nr}$ indicated by 
upper axis of the plot will be discussed in the next section.

%The S2/S1 ratio of ER events has a minimum around 30~$\rm p.e.$ due to recombination law 
%which is well-known in LXe case~\cite{shutt2007performance, dahl2009physics}. 
%Namely ER events follow Thomas-Imel Box (TIB) model~\cite{thomas1987recombination} in the lower energy region and 
%Doke-Birks model~\cite{doke1988let} in the higher energy region while NR events follow only TIB model resulted in monotonic decrease in lower energy region.
%
For ER events, the S2/S1 ratio has a minimum around S1 $\sim$ 30~p.e. as shown in Fig.~\ref{img:S1-LogS2S1_1kV/cm} (top).
This structure has been also observed in the LXe experiments~\cite{shutt2007performance, dahl2009physics}, 
and is explained by the difference in the recombination mechanism for events below and above the minimum.
When the ER events have smaller recoil energy and hence short tracks (typically shorter than the electron diffusion length),
electron-ion pairs are concentrated in a small sphere and they cause ``box recombination'' as described  
by the Thomas--Imel Box (TIB) model~\cite{thomas1987recombination}. 
In this case, recombination probability becomes larger for larger energy, then the S2/S1 ratio decreases.
Whereas, when the recoil electrons have larger energy and longer tracks, electron-ion pairs are distributed in a pillar shape 
and cause ``columnar recombination'' as described by the Doke--Birks model~\cite{doke1988let}. 
In this case, recombination probability becomes smaller for larger energy (with small $dE/dx$), then the S2/S1 ratio increases.
For NR events, the tracks are short in the energy from keV to several MeV, hence they are always described by the TIB model 
and the S2/S1 ratio decreases monotonically as S1 increases.

\begin{figure}[h]
\centering
\includegraphics[width=8cm,clip]{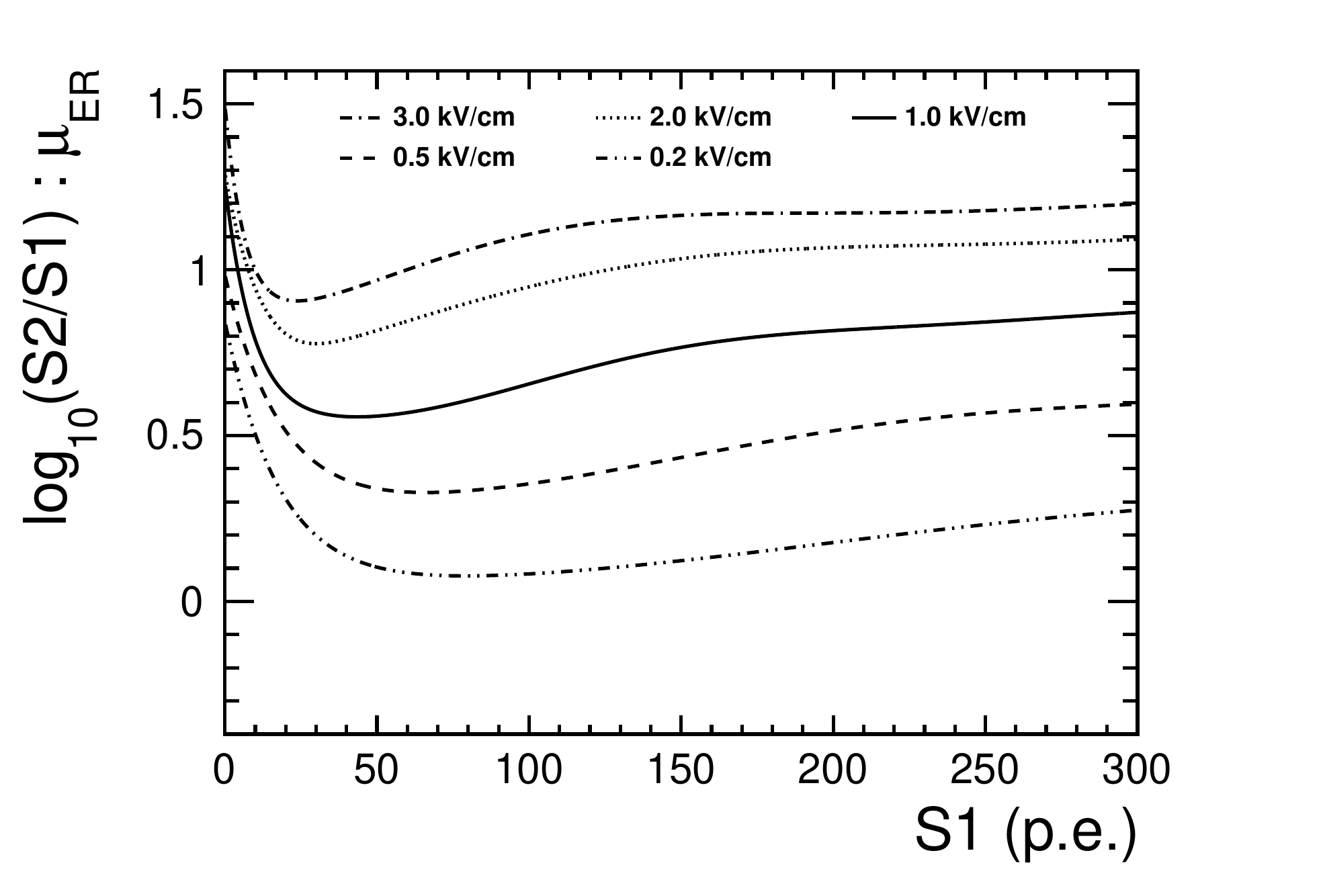}
\includegraphics[width=8cm,clip]{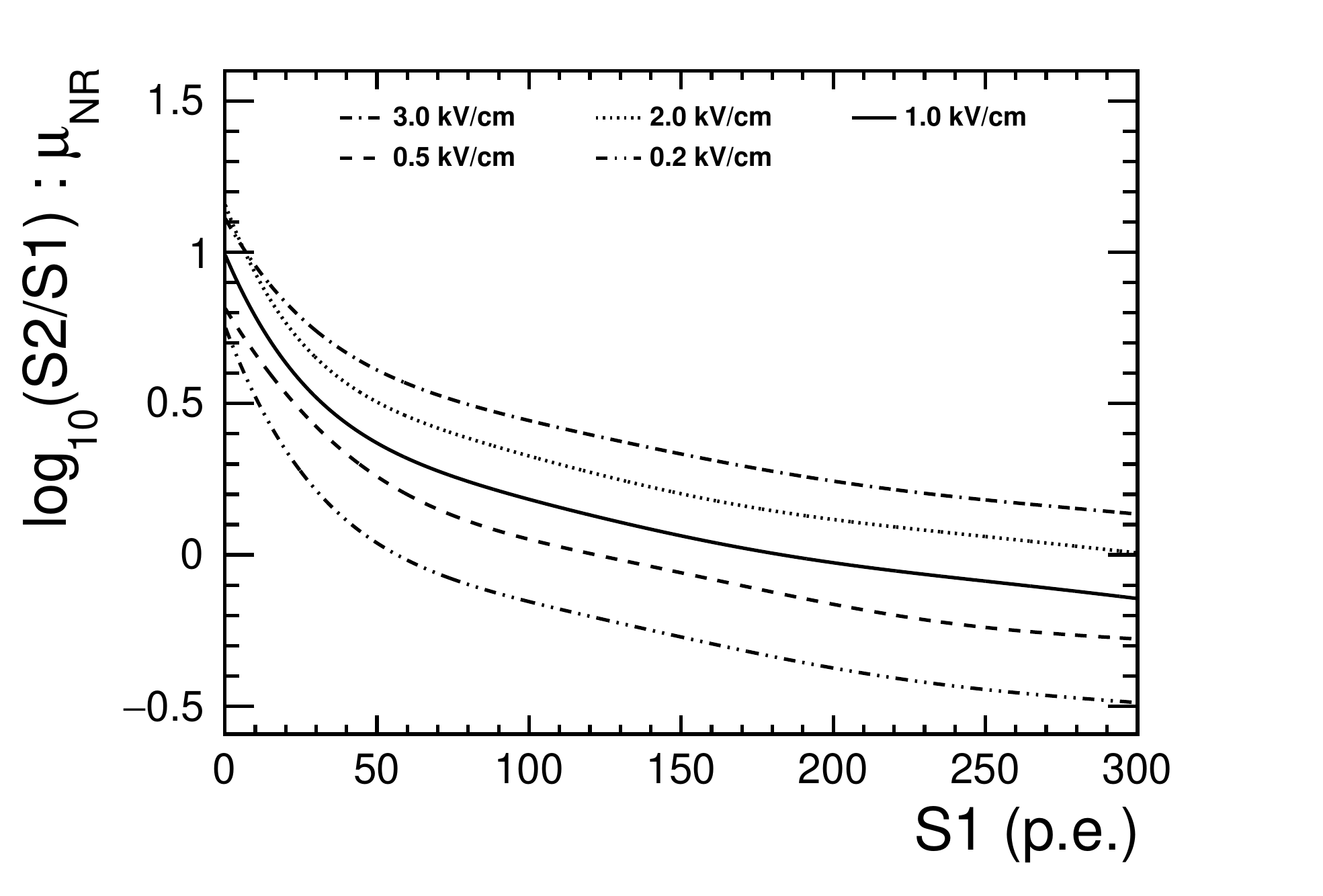}
\caption{The mean value, $\mu$ of  $\rm \log_{10}(S2/S1)$, as a function of S1 for each electric field. The top plot for ER events and 
the bottom for NR events.}
\label{img:LogS2S1mean}
\end{figure}

\begin{figure}[h]
\centering
\includegraphics[width=8cm,clip]{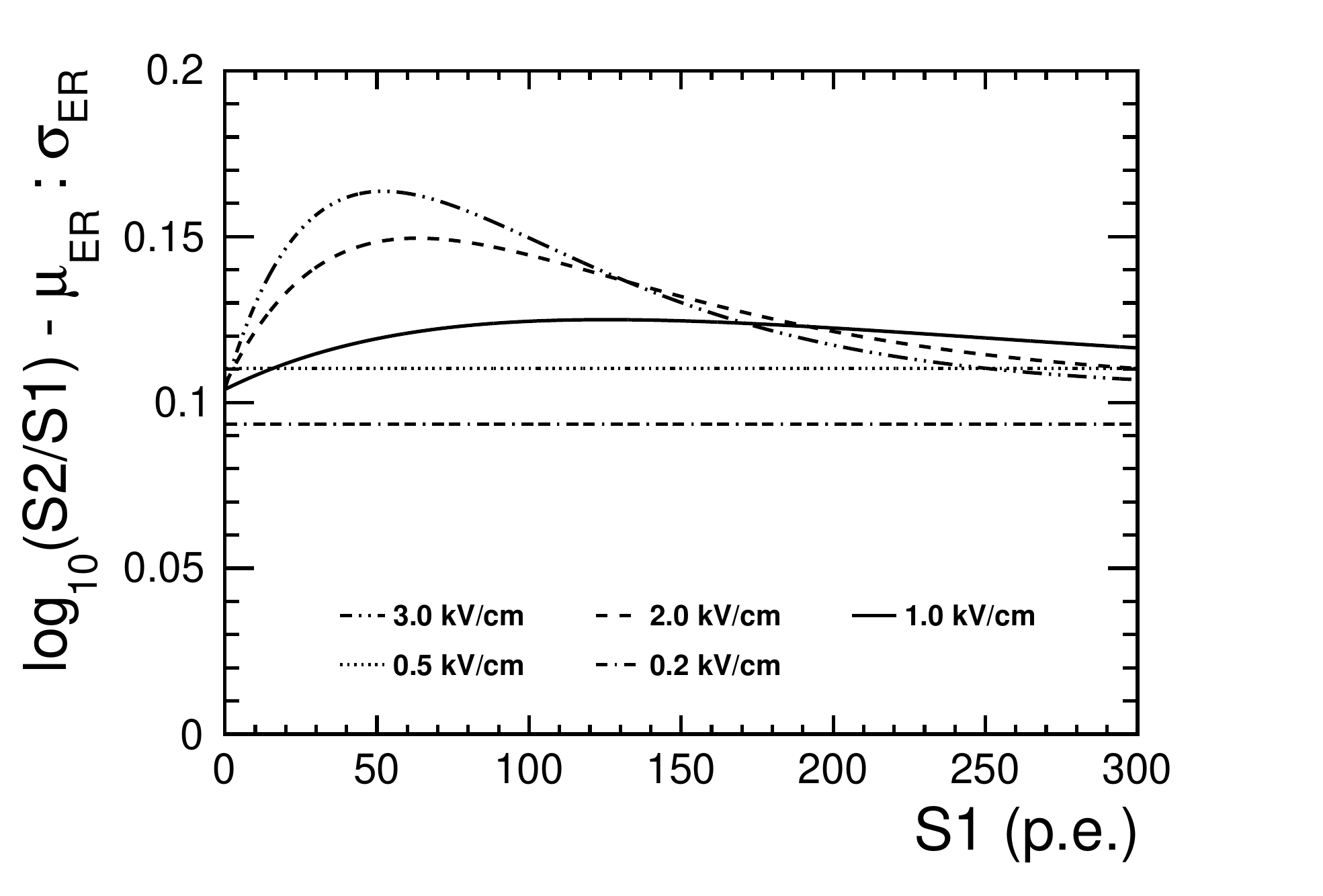}
\caption{The standard deviation, $\sigma_{_{\rm ER}}$ of  $\rm \log_{10}(S2/S1)-\mu_{_{\rm ER}}$, as a function of S1 for each electric field.}
\label{img:ERsigma}
\end{figure}

The same measurements and procedures are performed for various drift-fields,  0.2, 0.5, 1.0, 2.0, 3.0~kV/cm.
Energy dependence of the mean values, $\mu_{_{\rm ER}}$ and $\mu_{_{\rm NR}}$ at each electric field is shown in Fig.~\ref{img:LogS2S1mean}. 
As the electric field becomes higher, since recombination probability decreases, more S2 light yield  is observed compared to S1 light yield.
The standard deviations, $\sigma_{_{\rm ER}}$, from Gaussian fitting to ER events are summarized in Fig.~\ref{img:ERsigma},
while the one for NR events ($\sigma_{_{\rm NR}}$) is flat at 0.06, not depending on S1 nor drift-field.

%%%%%%%%%%%%%%%%%%%%%%%%%%%%%%%%%%%%%%%%%%%%%%%%%%%%%%%%%%%%%%%%%%%%%%%%%%%%
%\newpage
\section{Recoil energy and recombination law}
\label{sec:Recombination}
In order to evaluate the ER/NR discrimination power and its dependences of energy and electric field,
we need to convert S1 light yield to nuclear recoil energy $E_{\rm nr}$.
In this paper, the quenching factor measured by SCENE~\cite{cao2015measurement} below 1~kV/cm is extrapolated up to 3~kV/cm.

\begin{figure}[htbp]
\centering
\includegraphics[width=8cm,clip]{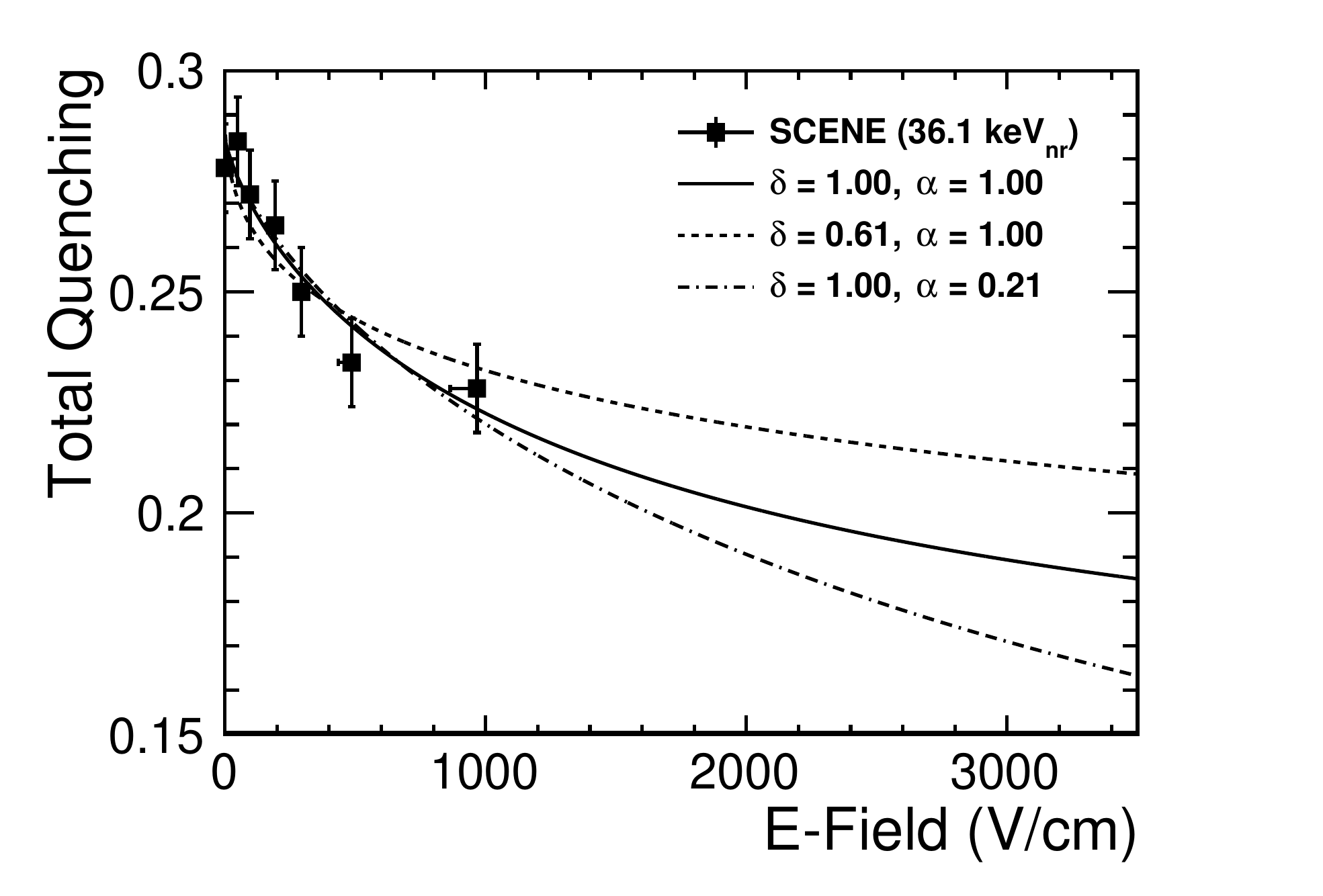}
\caption{Field dependence of the total quenching ($\mathcal{L}_{\rm eff}\times(\alpha+R)/(\alpha+1)$) 
measured by SCENE~\cite{cao2015measurement}  at $\rm 36.1~keV_{nr}$ and its  extrapolation (see text).}
\label{img:SCENEandFit}
\end{figure}

Fig.~\ref{img:SCENEandFit} shows the drift-field dependence of the total quenching including nuclear- and electric-quenching 
for S1 light yield measured by SCENE~\cite{cao2015measurement} at $\rm 36.1~keV_{nr}$ where the data points are only available up to 1~kV/cm. 
Extrapolation for higher electric field is performed by taking into account recombination law.

The S1 light yield can be expressed as a function of recoil energy $E_{\rm nr}$,
\begin{eqnarray}
{\rm S1} = LY \cdot E_{\rm nr} \cdot \mathcal{L}_{\rm eff} \cdot \frac{\alpha + R}{\alpha +1},  \label{eq:S1}
\end{eqnarray}
where $LY = 5.7 \ {\rm p.e./keV_{ee}}$ is the light yield for ER at null electric field, 
$\mathcal{L}_{\rm eff}$ is the nuclear quenching factor,
$\alpha = N_{\rm ex} / N_{\rm i}$ is the initial excitation/ionization ratio, 
and $R$ is the electron-ion recombination probability. 
Thus the electric quenching factor is given by $(\alpha+R)/(\alpha+1)$ in this formula~\cite{agnes2018measurement}.
For NR,  $\alpha$ is set to be unit as a priori input as done in~\cite{agnes2018measurement, agnes2017simulation}.

The nuclear quenching factor $\mathcal{L}_{\rm eff} = L\cdot f_l$ is written by the Mei model~\cite{mei2008model},
\begin{eqnarray}
L = \frac{k g (\epsilon)}{1+k g (\epsilon)}, \label{eq:Linhard} \\
f_l = \frac{1}{1+k_{\rm B} \frac{dE}{dx}}. \label{eq:Birks}
\end{eqnarray}
$L$ is the Lindhard factor~\cite{lindhard1963integral}, where
$\epsilon  = 11.5 E_{\rm nr} Z^{-7/3},$  
$g(\epsilon) = 3 \epsilon ^{0.15} + 0.7 \epsilon^{0.6} + \epsilon,$  
$k = 0.133 Z^{2/3} A^{-1.2},$  
with $E_{\rm nr}$ in keV and $Z, A$ as the atomic and mass numbers.
The factor $f_l$ explains the Birks saturation law, where $k_{\rm B} = 5.0\times 10^{-4}~{\rm MeV^{-1}~g~cm^{-2}}$~\cite{cao2015measurement}.

In the modified TIB model (c.f. in NEST~\cite{lenardo2015global} for LXe), $R$ is parametrized as follows, 
\begin{eqnarray}
R &=& 1- \frac{\ln (1+N_{\rm i }\varsigma )}{N_{\rm i }\varsigma},  \label{eq:R}  \\
\varsigma &=&  \gamma F^{-\delta},  \label{eq:varsigma} \\
N_{\rm i} &=& \frac{E_{\rm nr}}{W} \cdot \frac{1}{\alpha+1}\cdot \mathcal{L}_{\rm eff},  \label{eq:Ni} 
\end{eqnarray}
where $F$ is the drift-field, $N_{\rm i}$ is the number of ionizing electron, and $W = 19.5~{\rm eV}$~\cite{doke1988let,doke2002absolute} is the effective work function. 
In the original Tomas--Imel prediction,  $\delta$ is 1.0 which is consistent with the result 
of ARIS~\cite{agnes2018measurement}, while SCENE claims $\delta=0.61\pm0.03$ from the S2 behavior of $\rm ^{83m}Kr$ data.
In this paper, we employ $\delta=1.00$ and $\alpha=1.00$ as a baseline setup and the value $\gamma$ in Eq. (\ref{eq:varsigma}) is
derived from the fitting using all the data of SCENE (0--0.97~kV/cm, 10.3--57.3~$\rm keV_{nr}$), as shown in case 1 in Tab.~\ref{tab:FitResults}.
For other parameter settings, we compare case 2 ($\delta=0.61$) and case 3 ($\alpha=0.21$~\cite{doke1988let}) as 
a source of systematic uncertainty for the ER/NR discrimination power estimation described in the next section.

\begin{table}[htbp]
\centering
  \begin{tabular}{cccc}\hline\hline
                  &  	$\delta$      & $\alpha$ &	$\gamma\ [{\rm (V/cm)^\delta /e^-}]$ \\ \hline
      case 1   &  	1.00          &	1.00     &	$13.9\pm1.9$  \\
      case 2   &  	0.61          &	1.00     &	$1.2\pm0.2$ \\
      case 3   &       1.00          &	0.21     &	$35.7\pm3.9$  \\
      ARIS~\cite{agnes2018measurement}  &    $1.07\pm0.09$ &	1.00     & 	$18.5\pm9.7$    \\ \hline\hline
  \end{tabular}
\caption{Three cases of $\delta$ and $\alpha$ parameter setting and fitting results of $\gamma$ extracted 
by SCENE data with a comparison to the ARIS result~\cite{agnes2018measurement}.}
\label{tab:FitResults}
\end{table}

\begin{figure}[ht]
\centering
\includegraphics[width=8cm,clip]{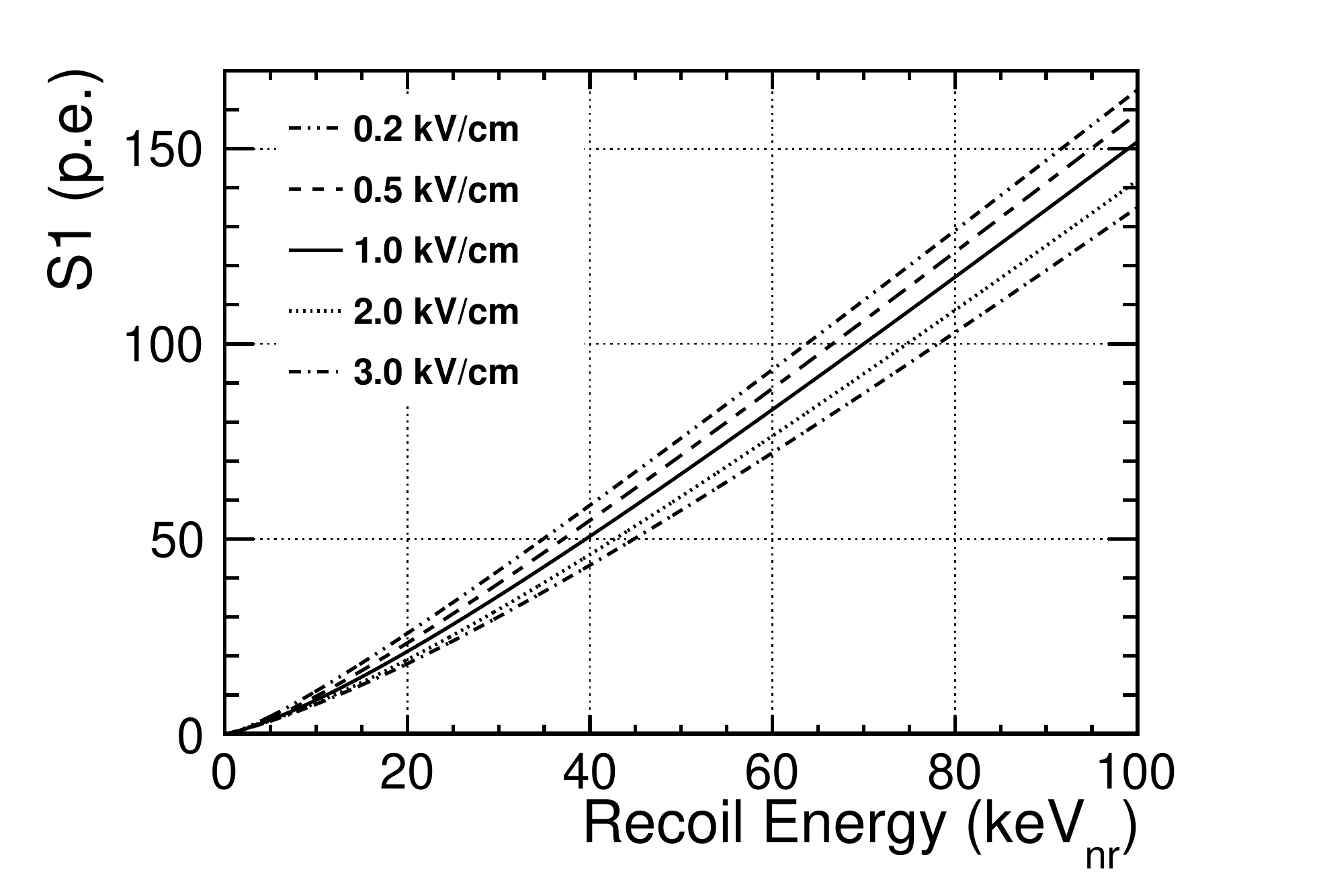}
\caption{Relation between S1 and recoil energy with $\delta=1.00, \alpha=1.00$ for each drift-field.}
\label{img:S1-keVnr}
\end{figure}

The relation between S1 and $E_{\rm nr}$ from Eq.~(\ref{eq:S1}) is shown in Fig.~\ref{img:S1-keVnr}, 
and the recoil energy indicated in Fig.~\ref{img:S1-LogS2S1_1kV/cm} is given by this function.

%%%%%%%%%%%%%%%%%%%%%%%%%%%%%%%%%%%%%%%%%%%%%%%%%%%%%%%%%%%%%%%%%%%%%%%%%%%%
%\newpage
\section{ER/NR events discrimination power}
\label{sec:Discrimination}
\begin{figure}[h]
\centering
\includegraphics[width=8cm,clip]{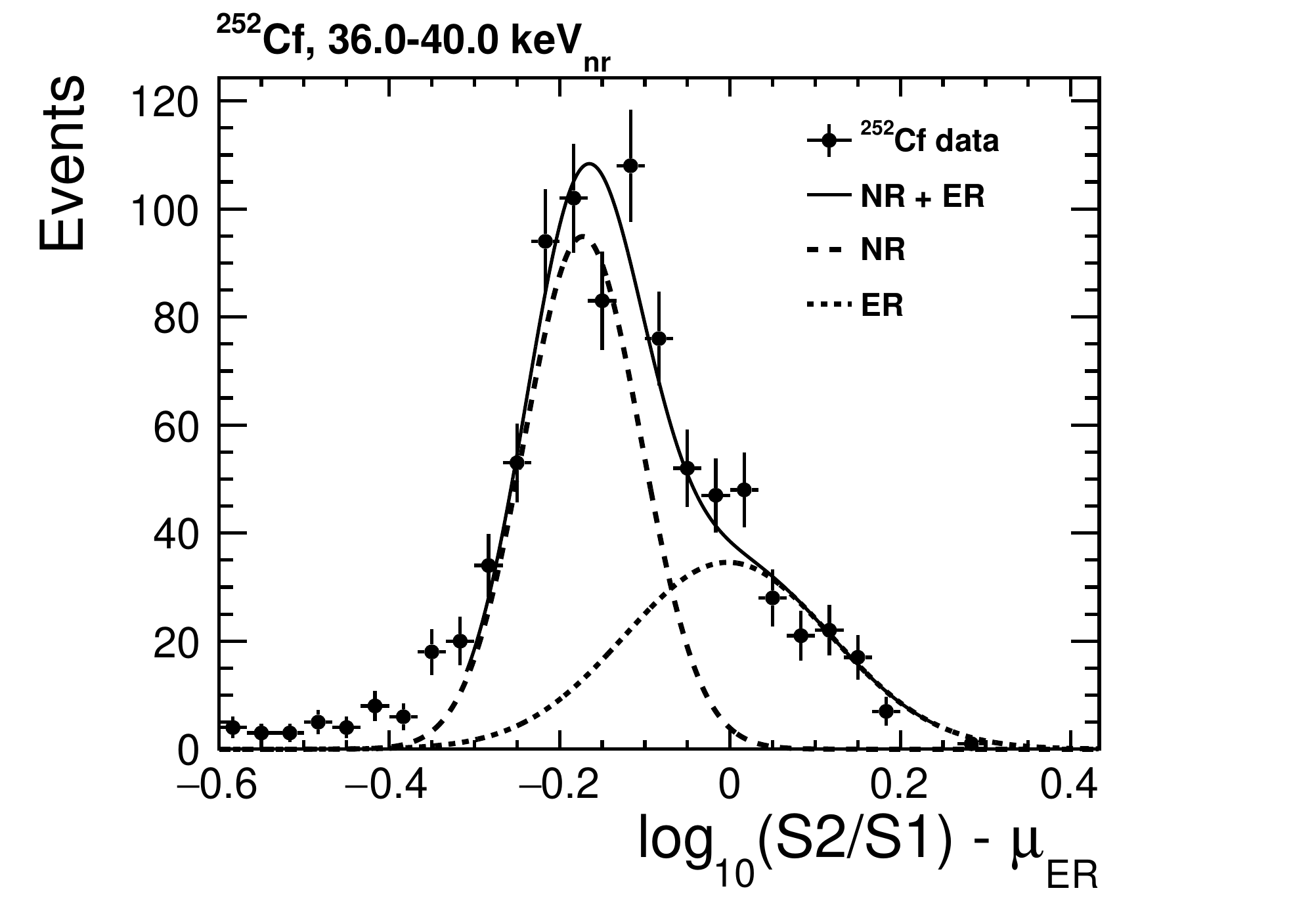}
\caption{ $\log_{10}({\rm S2/S1}) - \mu_{_{\rm ER}}$ distribution and two-Gaussian fitting of $\rm ^{252}Cf$ data in 36--40~$\rm keV_{nr}$ at 1.0~kV/cm.}
\label{img:36-40keVnr}
\end{figure}

The discrimination power between ER and NR is defined to be $(\mu_{_{\rm ER}}-\mu_{_{\rm NR}})/ \sigma_{_{\rm ER}}$.
After fitting the ER and NR peaks with two-Gaussian functions, the ER leakage fraction  to the NR signal region is
defined to be the ER fraction below the NR mean of $\mu_{_{\rm NR}}$.
For example, Fig.~\ref{img:36-40keVnr} shows the $\log_{10}({\rm S2/S1}) - \mu_{_{\rm ER}}$ distribution 
of the $\rm ^{252}Cf$ data within the recoil energy region of 36--40~$\rm keV_{nr}$ at 1.0~kV/cm.
As a result of two Gaussian fitting to determine $\mu_{_{\rm ER}}, \sigma_{_{\rm ER}}, \mu_{_{\rm NR}}$ and $\sigma_{_{\rm NR}}$, 
the discrimination power is calculated to be $(\mu_{_{\rm ER}}-\mu_{_{\rm NR}})/ \sigma_{_{\rm ER}}=1.40\pm0.06$. 
It is equivalent to the ER leakage fraction of $8.0\times 10^{-2}$.

\begin{figure}[h]
\centering
\includegraphics[width=9cm,clip]{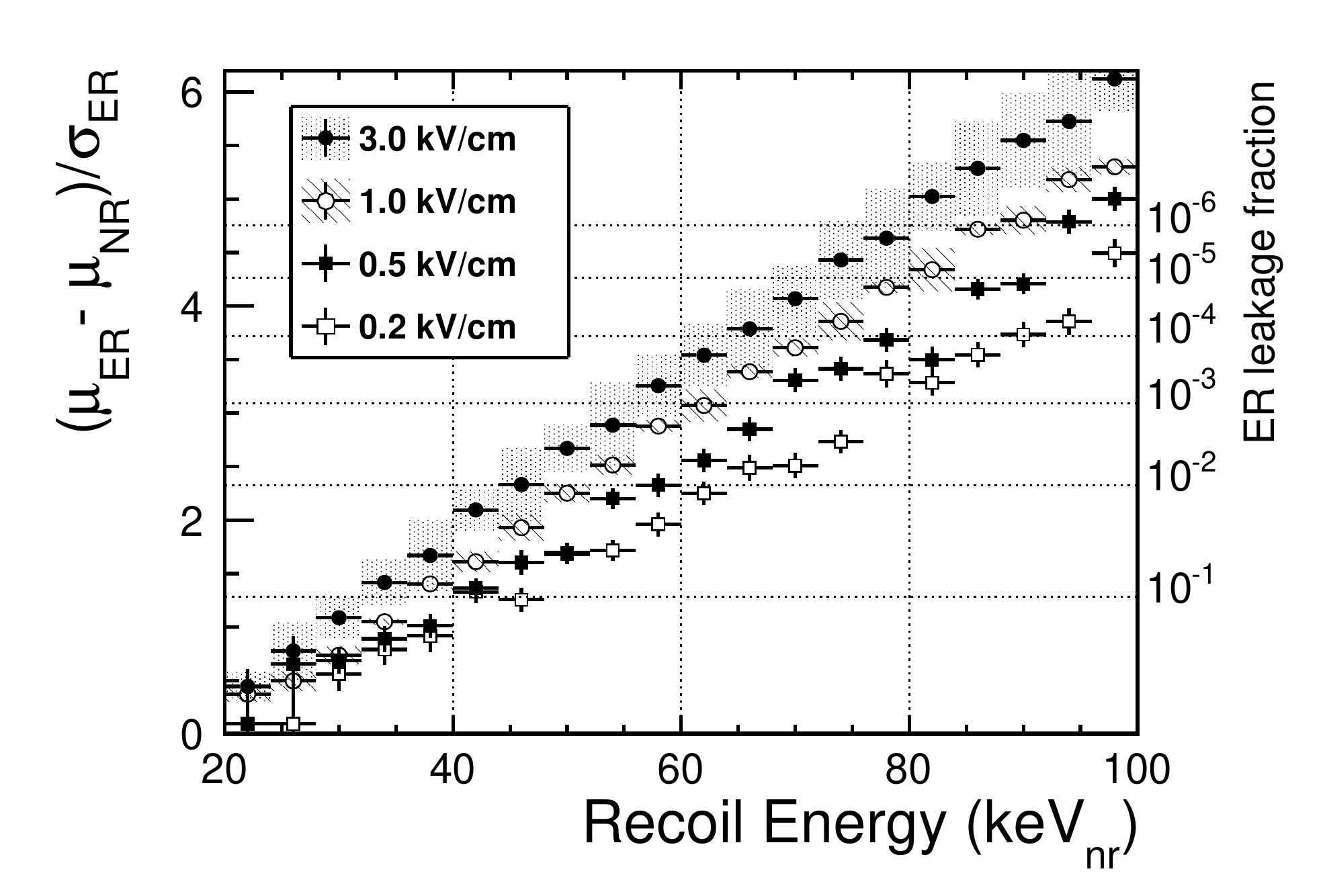}
\caption{Gaussian-extrapolated ER leakage fraction, at 50\% acceptance of NR, as a function of recoil energy for each drift-field. 
%(2~kV/cm result is removed from this figure to make these plots easy to see, but it exists between 1~kV/cm and 3~kV/cm results.)
}
\label{img:Rejection}
\end{figure}

The same fitting is performed for all the sets of drift-fields, within each recoil energy bin width of $\rm 4~keV_{nr}$  
and the results are summarized in Fig.~\ref{img:Rejection}. 
% 系統誤差の話。
For $F \geq 1~{\rm kV/cm}$ dataset,  $(\mu_{_{\rm ER}}-\mu_{_{\rm NR}})/ \sigma_{_{\rm ER}}$ is also calculated for the cases 1, 2, and 3 of the Tab.~\ref{tab:FitResults}, 
to take the uncertainty of the quenching model into account. 
In this region of $E_{\rm nr}$, 20--100~$\rm keV_{nr }$, the discrimination power becomes better as increasing energy for all drift-fields. 
When compared at the same recoil energy, higher field makes better discrimination.

%%%%%%%%%%%%%%%%%%%%%%%%%%%%%%%%%%%%%%%%%%%%%%%%%%%%%%%%%%%%%%%%%%%%%%%%%%%%
%\newpage
\section{Conclusion}
\label{sec:Conclusion}
We have reported the S2/S1 properties of a two-phase argon detector for both ER and NR events at drift-fields from 0.2~kV/cm to 3.0~kV/cm.
The discrimination power is improved at higher field in the recoil energy region of 20--100~$\rm keV_{nr}$. 
For the WIMP signal (NR event) search with argon, it is crucial to remove intrinsic ER background events caused by  $^{39}$Ar radio-isotope (about 1~Bq/kg in atmospheric argon).
Therefore, optimization of drift-field to maximize the ER rejection power for each experimental environment plays an important role for the physics sensitivity.
Our results would be useful for the design, operation and analysis of the current and future two-phase argon detector experiments for the WIMP search.

%%%%%%%%%%%%%%%%%%%%%%%%%%%%%%%%%%%%%%%%%%%%%%%%%%%%%%%%%%%%%%%%%%%%%%%%%%%%
\section*{Acknowledgments}
\label{sec:Acknowledgments}
% 早稲田特定課題
This work is a part of the outcome from research performed under the Waseda University Research Institute for Science and Engineering (Project numbers 2016A-507), 
% 新学術地下素核、
supported by the JSPS Grant-in-Aid for Scientific Research on Innovative Areas Grant Numbers 17H05204 and 15H01038,
%学振DC2
and the Grant-in-Aid for JSPS Research Fellow Grant Number 16J06656.



  \bibliographystyle{elsarticle-num} 
%  \bibliography{bib}

  \bibliography{bib_Title}
  %\bibliography{bib_Title2}


%\begin{thebibliography}{00}


%\bibitem{}

%\end{thebibliography}
\end{document}